%% file: barrier_sensing.tex
\documentclass[letterpaper, 10 pt, conference]{ieeeconf}  % Comment this line out if you need a4paper

\IEEEoverridecommandlockouts                              % This command is only needed if
\input{common_preamble.tex}

\newtheorem{theorem}{Theorem}
\newtheorem{lemma}{Lemma}
\newtheorem{corollary}{Corollary}
\theoremstyle{definition}
\newtheorem{example}{Example}
\newtheorem{assumption}{Assumption}

\theoremstyle{definition}
\newtheorem{definition}{Definition}
\theoremstyle{remark}
\newtheorem{remark}{Remark}

\title{\Large \bf
Safety With Limited Range Sensing Constraints For Fixed Wing Aircraft
}

\author{Eric Squires, Rohit Konda, Pietro Pierpaoli, Samuel Coogan, and Magnus Egerstedt$^\dagger$%
\thanks{$^\dagger$Eric Squires is with the Georgia Tech Research Institute.
    Pietro Pierpaoli, Samuel Coogan, and
    Magnus Egerstedt are with the School of Electrical and
    Computer Engineering,
    Georgia Institute of Technology, Atlanta, GA 30332, USA.
    Rohit Konda is with the University of California Santa Barbara, Santa Barbara, California 93106.
    The work of Eric Squires was supported by
    the University System of Georgia's Tuition Assistance Program.
    The work by Magnus Egerstedt and Pietro Pierpaoli was 
    supported by Grant No. ARL DCIST CRA W911NF-17-2-0181 by the US Army Research Lab.
    The work done by Samuel Coogan was supported by NSF Grant \#1924978.\looseness=-1
}
}

\begin{document}

\renewcommand{\onlyinsubfile}[1]{}
\renewcommand{\notinsubfile}[1]{#1}

\maketitle
\thispagestyle{empty}
\pagestyle{empty}

\input{abstract.tex}
\input{intro.tex}
\input{background.tex}

\input{motivation.tex}
\input{sensing.tex}
\input{alpha.tex}
\input{sim.tex}

\addtolength{\textheight}{-8.7cm}   % This command serves to balance the column lengths

\subfile{conclusion.tex}

\bibliographystyle{IEEEtran}
\bibliography{barrier.bib}

\end{document}

%% file: common_preamble.tex
\usepackage[hidelinks]{hyperref}
\usepackage{cite}
\usepackage{subfiles}
\newcommand{\onlyinsubfile}[1]{#1}
\newcommand{\notinsubfile}[1]{}

\usepackage[dvipsnames]{xcolor}
\usepackage[normalem]{ulem}

\usepackage{amsmath} % assumes amsmath package installed
\usepackage{amssymb}  % assumes amsmath package installed
\usepackage{todonotes}

\usepackage{amsthm}
\usepackage{graphicx}
\usepackage{subcaption}

\newcommand{\set}[2]{\{#1\: : \: #2\}}
\newcommand{\norm}[1]{\left\lVert #1\right\rVert}
\newcommand{\T}[1]{\widetilde{#1}}
\newcommand{\INF}{\inf_{\tau\in[0,\infty)}}

\newcommand{\pf}[2]{\frac{\partial #1}{\partial #2}}

\newcommand{\R}{\mathbb{R}}
\newcommand{\ra}{\rightarrow}

\DeclareMathOperator{\sign}{sgn}

%% file: abstract.tex
\begin{abstract}
In this paper we discuss how to use a barrier function that is subject to kinematic constraints and limited sensing in order to guarantee that 
fixed wing unmanned aerial vehicles (UAVs) will maintain safe distances from each other at all times despite
being subject to sensing constraints. Prior work has shown that a barrier function can be used to
guarantee safe system operation when the state can be sensed at all times.
However, we 
show that this construction does not guarantee safety when the UAVs are subject to limited range sensing.
To resolve this issue, we
introduce a method for constructing a new barrier function 
that accommodates limited sensing range from a previously existing barrier function that may not necessarily accommodate limited range sensing.
We show that, under appropriate conditions, the newly constructed barrier function ensures system safety even in the presence of limited range sensing.
We demonstrate the 
contribution of this paper in a simulated scenario of 20 fixed wing aircraft where
the vehicles are able to maintain safe distances from each other even though the vehicles are subject to limited range sensing.
\end{abstract}

%% file: intro.tex
\section{INTRODUCTION}
\label{sec_intro}

% Fixed-wing unmanned aerial vehicles (UAVs) have numerous important
% applications such as security and search and rescue \cite{cai2010brief,
% kopardekar2016unmanned}. At the same time, their safety is subject to a variety of
% constraints, such as dynamics, actuator limits, and sensing, that can complicate their
% deployment.  In particular,
Given the proliferation of fixed-wing unmanned aerial vehicles (UAVs) \cite{faa2020forecast},
an important deployment consideration is collision
avoidance where vehicles must be able to maintain safe distances \cite{kopardekar2016unmanned}.
Dynamic
constraints imply that vehicles must begin avoidance maneuvers well in advance
of a potential collision but if available sensing offers too short of a horizon to
begin those maneuvers then safety may not be maintained.  In this paper focusing on fixed-wing vehicles, we
show how to achieve collision avoidance while simultaneously taking into account dynamics,
actuator constraints, and limited range sensing.

There have been a variety of approaches to fixed-wing collision avoidance that
have modeled sensing and kinematic constraints including potential
fields \cite{mastellone2008formation, rodriguez2014decentralized, panyakeow2010decentralized}, POMDPs
\cite{temizer2010collision, wolf2011aircraft}, model predictive control \cite{di2015potential, defoort2009motion, shin2009nonlinear}, first
order lookahead approaches \cite{lalish2008decentralized, fiorini1998motion}, and optimal control \cite{tomlin1998conflict}. 
Similarly, barrier functions have been used 
in the context of limited sensing \cite{borrmann2015control,wang2016safety,wu2016safetyquad,xu2017correctness}
and allow for safety guarantees so that when the system starts safe it will remain safe for all future time.
In \cite{borrmann2015control} the
authors provide a minimum sensing radius to ensure a system of double
integrator robots maintain safe distances from each other.
Further, they
reformulate a Quadratic Program (QP) that only requires knowing the relative
position to other agents while still ensuring safety for both collaborative
and non-collaborative neighbors. In \cite{wang2016safety}, the authors provide a
decentralized strategy for collision avoidance that does not require knowing
neighbor barrier function parameters.
While
\cite{wu2016safetyquad} does not address multi-agent systems, it does consider
collision avoidance under limited range sensing for 3D quadrotors.
The authors design a sequential QP that translates position-based constraints into rotational commands to ensure safety. 
Sensing limitations can also be addressed with a disturbance to the system
dynamics. For instance, in \cite{xu2017correctness} the authors model road
curvature changes as a bounded disturbance to apply barrier
functions to adaptive cruise control and lane keeping.

The above examples require designing a specialized barrier function that satisfies
sensor constraints, where the additional consideration of the sensor can complicate the construction of a barrier function.
Thus, in this paper, we decouple the construction of a barrier function from the
sensor constraint. We do this by demonstrating how to adjust 
a barrier function that does not consider sensing limitations
into one that can still be used to guarantee safety when sensing limitations are present.
We build on the work of \cite{squires2019composition}
where it was shown how to ensure a system of $k$
UAVs maintain safe distances for all time while 
taking into account dynamic constraints. However, \cite{squires2019composition} did not consider 
limited range sensing.
Thus, in this paper we relax this limitation with the following
contributions.
First, we show that the previous barrier functions do not necessarily guarantee safety when the UAVs are subject to limited range sensing.
Second, 
we
introduce a method for constructing a new barrier function 
that accommodates limited sensing range.
Finally,
we conduct a simulation consisting of a scenario of 20 UAVs, where because of the proposed algorithm,
the vehicles are able to maintain safe distances from each other even though the vehicles are subject to limited range sensing.

This paper is organized as follows: Section~\ref{sec_background} provides the
necessary background for barrier functions.  Section~\ref{sec_motivation}
discusses issues that can arise when using a barrier function to ensure safety
when there is limited range sensing.  Section~\ref{sec_limited_sensing}
proposes a novel approach to derive a barrier function that can be used to make
safety guarantees despite UAVs' sensing limitations. 
Section~\ref{sec_alpha} analyzes the set of safe control inputs
when using the proposed approach.
Section~\ref{sec_simulation} shows a simulation with twenty simulated fixed-wing UAVs to
verify the theoretical developments of this paper.

%% file: background.tex
\section{BACKGROUND}
\label{sec_background}

In this paper, we study the problem of collision avoidance among a team of UAVs
with limited sensing range. For ease of exposition, we focus on collision
avoidance between two UAVs. Because collision avoidance is a pairwise constraint,
we can then apply the method developed in this paper to all pairwise combinations
of aircraft to ensure safety of the entire set of UAVs
without requiring that each vehicle can sense all others at all times.
We model fixed-wing UAVs similarly to \cite{squires2019composition}, where
the state of UAV $i$ ($i \in \{1,2\}$) is given by
$x_i = \begin{bmatrix} p_{i,x} & p_{i,y} & \theta_i & p_{i,z} \end{bmatrix}^T$
with dynamics
\begin{equation}
\dot{x}_i
=
\begin{bmatrix}
v_i\cos\theta_i &
v_i\sin\theta_i &
\omega_i &
\zeta_i
\end{bmatrix}^T,
\label{eq_fixed_wing_dynamics}
\end{equation}
where $v_i\in [v_{min},v_{max}]$ is the linear velocity, $\omega_i\in [-\omega_{max},\omega_{max}]$ is the turn rate,
and $\zeta_i \in [-\zeta_{max},\zeta_{max}]$ is the altitude rate.
This model is applicable in cases where the roll and pitch of the aircraft 
are small \cite{jackson2011controlling}. Similar models have been used in fixed-wing 
collision avoidance as in 
\cite{
    lalish2008decentralized,
    % mastellone2008formation,
    rodriguez2014decentralized,
    panyakeow2010decentralized,
    di2015potential,
    pallottino2007decentralized,
    krontiris2011using}.
The dynamics of two aircraft is then 
$\dot{x} = \begin{bmatrix} \dot{x}_1^T & \dot{x}_2^T \end{bmatrix}^T$.
The system (\ref{eq_fixed_wing_dynamics}) is a particular instance of a control affine system 
\begin{equation}
\dot{x} = f(x) + g(x)u
\label{eq_system}
\end{equation}
where $f$ and $g$ are locally Lipschitz, $x\in\R^n$, $u\in U\subset \R^m$, and we assume the
system has a unique solution for all $t \ge 0$ given a starting condition $x(0)$ and input function $u$. 

Assume there is a safe set $\mathcal{C}$ that is the superlevel set of a continuously differentiable function $h$ so that
the safe set is
\begin{IEEEeqnarray}{rCl}
\mathcal{C} &=& \{x\in \R^n \: : \: h(x) \ge 0\}.
\label{eq_safe}
\end{IEEEeqnarray}
$\mathcal{C}$ represents the set of aircraft states where
collision avoidance can be guaranteed.
In the following, $L_fh(x) = \pf{h(x)}{x}f(x)$ and $L_gh(x) = \pf{h(x)}{x}g(x)$ denote Lie derivatives.
\begin{definition}
\cite{ames2017control}
Given a set $\mathcal{C}\subset \R^n$ defined in
(\ref{eq_safe}) for a continuously
differentiable function $h:\R^n\ra\R$, the function
$h$ is called a \emph{zeroing control barrier function (ZCBF)}
defined on an open set $\mathcal{D}$ with $\mathcal{C}\subset \mathcal{D}\subset \R^n$,
if there exists a Lipschitz continuous extended class $\mathcal{K}$ function $\alpha$
such that
\begin{equation}
\sup_{u\in U}[L_fh(x) + L_gh(x)u + \alpha(h(x))] \ge 0 \quad \forall x \in \mathcal{D}.
\label{eq_zcbf_condition} 
\end{equation}
\label{def_zcbf}
\end{definition}
\vspace*{-0.2in}
\noindent The admissible control space,
$K(x) = \{u \in U\: : \: L_fh(x) + L_gh(x)u + \alpha(h(x)) \ge 0\}$,
can be used to find sufficient conditions for the forward
invariance of $\mathcal{C}$, meaning that if $x(0)\in \mathcal{C}$
then $x(t)\in\mathcal{C}$ for all $t \ge 0$
where $x(t)$ is the solution to the closed loop system under a fixed controller.
\begin{theorem}\cite{ames2017control}
Given a set $\mathcal{C}\subset\R^n$ defined in (\ref{eq_safe})
for a continuously differentiable function $h$, if $h$ is a ZCBF on $\mathcal{D}$,
then any Lipschitz continuous controller $u:\mathcal{D}\ra U$ such that
$u(x)\in K(x)$ will render the set $\mathcal{C}$ forward invariant.
\label{th_fwd_invariance}
\end{theorem}

In addition to being able to make system safety guarantees, a ZCBF can
be used to calculate a safe control input in an online manner through a
QP. Given a nominal control input $\hat{u}$
and a set of  actuator constraints that can
be expressed as linear inequalities, $Au \ge b$, we can calculate a safe
control input closest in norm to $\hat{u}$ as follows:
\begin{IEEEeqnarray}{Cl}
\label{eq_general_qp}
\min_{u\in \R^m} & \frac{1}{2}\norm{u - \hat{u}}^2 \nonumber\\
\text{s.t.} & L_fh(x) + L_gh(x) u + \alpha(h(x)) \ge 0 \nonumber\\
& Au \ge b.
\end{IEEEeqnarray}

The application of Theorem~\ref{th_fwd_invariance}
is conditional upon the definition of a suitable barrier function
that captures the safety requirement.
This can
be difficult when the system is subject to actuator constraints and nonlinear dynamics,
as is the case for fixed-wing UAVs.
Motivated by this, in \cite{squires2019composition}, it was shown how
to systematically construct a barrier function for fixed-wing UAVs.
Although the formulation in \cite{squires2019composition} does
not consider sensing limitations, we will show in this paper that 
the formulation can be adapted to create a barrier function that allows for
safety guarantees even when there are sensing restrictions.
In particular, let $\rho:\mathcal{D}\to \R$ be a safety function that
must be nonnegative at all times for the system to be considered safe.
Let $\gamma:\mathcal{D}\to U$ be a nominal evading maneuver.
Then a barrier function can be defined as the worst
case safety function value when using $\gamma$
for all future time,\looseness=-1
\begin{IEEEeqnarray}{rCl}
h(x;\rho,\gamma)
&=& \INF \rho(\hat{x}(\tau))
\label{eq_h}
\end{IEEEeqnarray}
where
$\dot{\hat{x}}(\tau) = f(\hat{x}(\tau)) +
g(\hat{x}(\tau))\gamma(\hat{x}(\tau))$
and
\begin{equation}
\hat{x}(\tau) = x(t) + \int_0^\tau \dot{\hat{x}}(\eta)d\eta.
\label{eq_h_int}
\end{equation}

\begin{theorem} \cite{squires2019composition}
Given a dynamical system (\ref{eq_system})
and a set $\mathcal{C}\subset \mathcal{D}$ defined in (\ref{eq_safe}) for
a continuously differentiable $h$ defined in (\ref{eq_h})
with a safety function $\rho$ and locally Lipschitz evading maneuver $\gamma$,
$h$ satisfies (\ref{eq_zcbf_condition}) for all $x\in \mathcal{C}$.
If in addition, $L_gh(x)$ is non-zero for all $x\in \partial \mathcal{C}$ 
and $\gamma$ maps to values in the interior of $U$,
then $h$ is a ZCBF on an open set $\mathcal{D}$ where $\mathcal{C}\subset \mathcal{D}$.
\label{th_h_zcbf} 
\end{theorem}

In \cite{squires2019composition}, two cases of $\gamma$ and $\rho$ are
considered for the calculation of 
$h$ in (\ref{eq_h}) in closed form.
First
\begin{equation}
\gamma_{turn} = \begin{bmatrix} \sigma v & \omega & 0 & v & \omega & 0 \end{bmatrix}^T
\label{eq_gamma_turn}
\end{equation}
is an evasive maneuver encoding a constant rate turn for both vehicles
with possibly different forward velocity where $0< \sigma \le 1$. The safety function
is
\begin{equation}
\rho_{turn}(x) = 
\sqrt{d_{1,2}(x) - 2\delta + \delta \sin(\theta_{1}) - \delta \cos(\theta_1)} - D_s,
\label{eq_rho_adjusted_sqrt}
\end{equation}
where
$\delta > 0$ is a scalar,
$d_{1,2}(x)$ is the squared distance between vehicles $1$ and $2$,
and $D_s$ is the minimum distance between the vehicles for the system
to be considered safe.
The second case considered in \cite{squires2019composition} uses
\begin{equation}
\gamma_{straight} = \begin{bmatrix} v_1 & 0 & \zeta_1 & v_2 & 0 & \zeta_2 \end{bmatrix}^T,
\label{eq_gamma_straight}
\end{equation}
where $v_1 \ne v_2$,
which encodes each vehicle maintaining a straight trajectory.
The safety function is
\begin{equation}
\rho_{straight}(x(t)) = \sqrt{d_{1,2}(x)} - D_s.
\label{eq_rho}
\end{equation}
When $h$ is constructed from $\gamma_{straight}$ and $\rho_{straight}$
we denote the resulting $h$ in equation (\ref{eq_h}) by $h_{straight}$.
We similarly denote $h_{turn}$ when constructing $h$ from $\gamma_{turn}$ and $\rho_{turn}$.
Motivated by the existence of a closed form solution to (\ref{eq_h}) for $h_{turn}$ and $h_{straight}$
we will consider $h_{turn}$ and $h_{straight}$ throughout this paper. However,
this particular choice is not central to the contribution of this paper.
% When $k>2$ there are $q = (k(k-1))/2$ pairwise distance safety constraints and
% an additional assumption is needed to ensure safety for all pairwise safety
% constraints. In the case of $q$ constraints, there are $q$ barrier functions
% which will be denoted by $h^j$ for $j=1,\ldots,q$.
% \begin{definition}
% Suppose every $h^j$ ($j=1,\ldots,q$) is defined as in (\ref{eq_h})
% and is constructed from locally Lipschitz $\gamma^j$, respectively.
% The
% \emph{shared evading maneuver assumption} holds
% if 
% $\gamma^1(x) = \cdots = \gamma^q(x)$
% for all $x\in \mathcal{D}$.
% The \emph{shared evading maneuver} is denoted $\gamma^s$ 
% so that $\gamma^s(x) = \gamma^1(x) = \cdots = \gamma^q(x)$
% for all $x\in \mathcal{D}$.
% \end{definition}

%% file: motivation.tex
\section{MOTIVATION}
\label{sec_motivation}

We assume there is a sensor modeled via a set $S\subset \mathcal{D}$ such that,
if the system state $x$ is such that $x\in S$, then $x$ is completely known to
both vehicles, whereas if $x\not\in S$, then all that is known is that
$x\not\in S$. This is the case for instance when the sensor has limited
range, as was considered in \cite{borrmann2015control,wu2016safetyquad}.
In the case of UAV collision avoidance where each UAV is equipped with an omnidirectional sensor with range $R$,
$S = \set{x\in \mathcal{D}}{d_{1,2}(x) \le R^2}$.
In this section we present two motivating examples to illustrate two distinct
issues that can arise when using barrier functions in the presence of limited
range sensing. In both cases, the critical problem is that $K(x)$
cannot be calculated for all $x\in\mathcal{D}$ because $S\subset \mathcal{D}$.

\begin{figure}
\centering
\begin{subfigure}[b]{0.45\columnwidth}
\centering
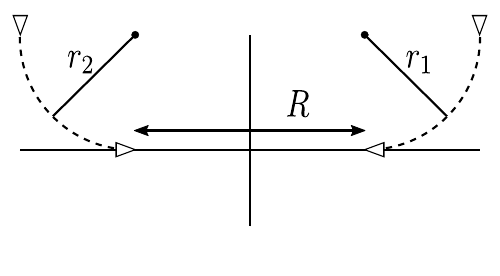
\caption{ }
\label{fig_examples_1}
\end{subfigure}
\qquad
\begin{subfigure}[b]{0.45\columnwidth}
\centering
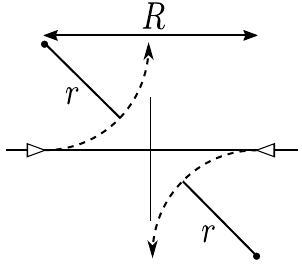
\caption{ }
\label{fig_examples_2}
\end{subfigure}
\caption{
Two examples where limited range sensing creates issues when applying barrier certificates to fixed-wing UAV collision avoidance.
In (a), the vehicles start so that
$h_{straight}(x) = (r_1 + r_2 + R) - D_s \ge 0$ but because the vehicles cannot sense
each other, achieve a configuration where $h_{straight}(x) = -D_s$.
In (b), the vehicles travel along the $x$-axis until they sense each other at a distance of $R$
apart, at which point the safe control implied by $h_{turn}$ requires high turn speed.
\vspace*{-0.25in}
}
\label{fig_examples}
\end{figure}

The first issue that limited range sensing introduces is that safety can no longer be guaranteed. In
particular, we construct a scenario where $h(x(0)) \ge 0$ and, because
$K(x)$ cannot be calculated, there is a future time for which $h(x(t)) < 0$.  In
other words, we can have a continuously differentiable barrier function $h$
that satisfies (\ref{eq_zcbf_condition}) but still not be able to guarantee
safety.  The second issue we examine is that there can be discontinuities in actuator
commands even though $h$ is continuously differentiable.  This can cause alarm
or discomfort for systems designed to ensure safety of human passengers (e.g.,
cruise control \cite{ames2017control}).  In the following examples, consider two UAVs equipped with omnidirectional sensors (e.g.
radar) of radius $R$, with dynamics governed by a nominal controller
$\hat{u}(x)$. See Fig. \ref{fig_examples} for illustrations.

\begin{example}
\label{ex_zcbf_starts_safe_not_safe}
\emph{A Barrier Function Without a Safety Guarantee.}
Suppose the two vehicles start at
$
x_1(0)= \begin{bmatrix} r_1 + R / 2 & r_1 & -\pi/2 & 0 \end{bmatrix}^T,
x_2(0)= \begin{bmatrix} -r_2 -R / 2 & r_2 & -\pi/2  & 0\end{bmatrix}^T$,
respectively, where 
vehicle 1 has a nominal control input of $\hat{u}_1 = \begin{bmatrix} v_1 & -\omega \end{bmatrix}^T$
and vehicle 2 has a nominal control input of $\hat{u}_2 = \begin{bmatrix} v_2 & \omega \end{bmatrix}^T$
so that they both follow a circular trajectory with radius $r_1 = v_1/\omega$ and $r_2 / \omega$, respectively (see Fig. \ref{fig_examples_1}).
Then $h_{straight}(x(0)) = (r_1 + r_2 + R) - D_s \ge 0$ as long as
$R\ge \max(0, D_s - r_1 - r_2)$ so the vehicles start safe
according to $h_{straight}$.
Further, note that because the vehicles cannot sense each other,
$K(x)$ cannot be calculated. This is because to calculate
$K(x)$, the values of $L_fh(x)$, $L_gh(x)$, and $h(x)$
are required.
Because $K(x)$ cannot be calculated,
there is no means to ensure that the control input
applied to the vehicle will be in $K(x)$.
In particular, it means that it is unknown
whether the nominal control input $\hat{u}$
is in $K(x)$ and a design decision must be employed for what actuation input to apply to the vehicles.
If the design decision is, for instance, to apply the nominal controller to the vehicles
then the vehicles will reach
$\begin{bmatrix} R / 2 & 0 & \pi & 0 \end{bmatrix}^T$
and 
$\begin{bmatrix} -R / 2 & 0 & 0  & 0\end{bmatrix}^T$,
respectively.
Once the vehicles have reached
this state, $h_{straight}(x) = -D_s$.
This means that the vehicles started in a state such that $h(x(0)) \ge 0$
but there exists a later time $t$ such that $h(x(t)) < 0$.
This is because $K(x)$ cannot be calculated for $x\notin S$
so the control input applied to the aircraft does not always satisfy (\ref{eq_zcbf_condition}).
In other words, because
$K(x)$ cannot be calculated
for all $x\in\mathcal{D}$,
Theorem \ref{th_fwd_invariance} cannot be used to guarantee safety.
\end{example}

\begin{example}
\label{ex_discontinuity}
\emph{Loss of Smoothness.}
For $h_{turn}$ let $\gamma_{turn}$ be specified
with $v = v_{min}$ and $\omega = \omega_{max}$ in (\ref{eq_gamma_turn}).
Let the two aircraft have sensor radius $R = (D_s + 2r)\cos(\eta) + 4\delta$
where $r = v_{min} / \omega_{max}$ and $\eta = \arcsin(r / (r + D_s / 2))$.
As in Fig.~\ref{fig_examples_2}, let the vehicles have initial positions of 
$\begin{bmatrix} (D_s/2 + r)\cos(\eta) + 2\delta + \epsilon & 0 & -\pi & 0 \end{bmatrix}^T$
and 
$\begin{bmatrix} -(D_s/2 + r)\cos(\eta) - 2\delta - \epsilon & 0 & 0 & 0\end{bmatrix}^T$,
respectively, where $\epsilon > 0$.
Further, let each aircraft have a nominal trajectory that continues toward
the origin. Because the aircraft cannot sense each other, 
$K(x)$ cannot be calculated.
This means that there
will be no collision avoidance override so the applied actuator command will be equal to 
the nominal controller command of $\hat{u}_i(x) = \begin{bmatrix} v_{max} & 0 \end{bmatrix}^T$
until the vehicles reach states
$\begin{bmatrix} (D_s/2 + r)\cos(\eta) + 2\delta & 0 & -\pi & 0 \end{bmatrix}^T$
and 
$\begin{bmatrix} -(D_s/2 + r)\cos(\eta) - 2\delta & 0 & 0  & 0\end{bmatrix}^T$,
respectively.
At this point, the vehicles can sense each other and the constraints (\ref{eq_zcbf_condition}) in the QP (\ref{eq_general_qp})
can be calculated, resulting
in a discontinuity in the 
constraint in the QP (\ref{eq_general_qp}).
\end{example}

%% file: img/svg/ex1.pdf_tex
%% Accompanies image file 'img/svg/ex1.pdf' (pdf, eps, ps)
%%   \input{<filename>img/svg/.pdf_tex}
%%   \includegraphics{<filename>img/svg/.pdf}
%%   \input{<filename>img/svg/.pdf_tex}
%%   \includegraphics[width=<desired width>]{<filename>img/svg/.pdf}
%%   \import{<path to file>}{<filename>img/svg/.pdf_tex}
\begingroup%
  \makeatletter%
  \providecommand\color[2][]{%
    \errmessage{(Inkscape) Color is used for the text in Inkscape, but the package 'color.sty' is not loaded}%
    \renewcommand\color[2][]{}%
  }%
  \providecommand\transparent[1]{%
    \errmessage{(Inkscape) Transparency is used (non-zero) for the text in Inkscape, but the package 'transparent.sty' is not loaded}%
    \renewcommand\transparent[1]{}%
  }%
  \providecommand\rotatebox[2]{#2}%
  \newcommand*\fsize{\dimexpr\f@size pt\relax}%
  \newcommand*\lineheight[1]{\fontsize{\fsize}{#1\fsize}\selectfont}%
  \ifx\svgwidth\undefined%
    \setlength{\unitlength}{143.8568728bp}%
    \ifx\svgscale\undefined%
      \relax%
    \else%
      \setlength{\unitlength}{\unitlength * \real{\svgscale}}%
    \fi%
  \else%
    \setlength{\unitlength}{\svgwidth}%
  \fi%
  \global\let\svgwidth\undefined%
  \global\let\svgscale\undefined%
  \makeatother%
  \begin{picture}(1,0.5508583)%
    \lineheight{1}%
    \setlength\tabcolsep{0pt}%
    \put(0,0){\includegraphics[width=\unitlength,page=1]{img/svg/ex1.pdf}}%
  \end{picture}%
\endgroup%

%% file: img/svg/ex2.pdf_tex
%% Accompanies image file 'img/svg/ex2.pdf' (pdf, eps, ps)
%%   \input{<filename>img/svg/.pdf_tex}
%%   \includegraphics{<filename>img/svg/.pdf}
%%   \input{<filename>img/svg/.pdf_tex}
%%   \includegraphics[width=<desired width>]{<filename>img/svg/.pdf}
%%   \import{<path to file>}{<filename>img/svg/.pdf_tex}
\begingroup%
  \makeatletter%
  \providecommand\color[2][]{%
    \errmessage{(Inkscape) Color is used for the text in Inkscape, but the package 'color.sty' is not loaded}%
    \renewcommand\color[2][]{}%
  }%
  \providecommand\transparent[1]{%
    \errmessage{(Inkscape) Transparency is used (non-zero) for the text in Inkscape, but the package 'transparent.sty' is not loaded}%
    \renewcommand\transparent[1]{}%
  }%
  \providecommand\rotatebox[2]{#2}%
  \newcommand*\fsize{\dimexpr\f@size pt\relax}%
  \newcommand*\lineheight[1]{\fontsize{\fsize}{#1\fsize}\selectfont}%
  \ifx\svgwidth\undefined%
    \setlength{\unitlength}{86.84504772bp}%
    \ifx\svgscale\undefined%
      \relax%
    \else%
      \setlength{\unitlength}{\unitlength * \real{\svgscale}}%
    \fi%
  \else%
    \setlength{\unitlength}{\svgwidth}%
  \fi%
  \global\let\svgwidth\undefined%
  \global\let\svgscale\undefined%
  \makeatother%
  \begin{picture}(1,0.91450731)%
    \lineheight{1}%
    \setlength\tabcolsep{0pt}%
    \put(0,0){\includegraphics[width=\unitlength,page=1]{img/svg/ex2.pdf}}%
  \end{picture}%
\endgroup%

%% file: sensing.tex
\section{CONSTRUCTING A BARRIER FUNCTION FOR SAFETY GUARANTEES DESPITE LIMITED RANGE SENSING RESTRICTIONS}
\label{sec_limited_sensing}

In Section~\ref{sec_motivation} we saw that limited range sensing can lead to
practical issues including the loss of safety guarantees even when
a ZCBF exists for the system. 
This means that UAVs may collide with each other when they are equipped with limited range sensors.
When limited sensing is not taken into account in the design of a ZCBF 
$h$,
the problem
is that values of $h$ cannot be
evaluated for all $x\in \mathcal{D}$, as required by Definition~\ref{def_zcbf},
and so $h$ cannot be used to guarantee safety. In this section we provide a
solution to this issue.

\begin{definition}
For a given ZCBF $h$ and a sensor with sensor set $S$, $h$ is \emph{sensor compatible}
if $h$ is a positive constant for all $x\notin S$.
\end{definition}
\begin{remark}
A sensor compatible 
ZCBF $h$ must be positive outside $S$
since otherwise 
this would imply for $x\notin S$
that (\ref{eq_zcbf_condition}) becomes
$\alpha(h(x)) < 0$ so (\ref{eq_zcbf_condition})
does not hold for any $u\in U$.
\end{remark}
\begin{remark}
For the case of UAVs equipped with a limited range sensor, this
means the value of $h$ is a positive constant when the vehicles are outside of the sensing range.
\end{remark}

Importantly, implementing a safety overriding controller requires an exact
calculation of the ZCBF constraint (\ref{eq_zcbf_condition}) only if $K(x) \ne U$.
When $K(x) = U$, there is no need to calculate $h(x)$ or its derivatives because
any $u\in U$ is already known to be safe.
Because of the additional structure on a sensor compatible ZCBF,
we can relax the need to check $u\in K(x)$ for all $x\in\mathcal{D}$
in Theorem~\ref{th_fwd_invariance}
because it is already known that $u\in K(x)$ for all $u$ when $x\notin S$.

\begin{corollary}
Suppose $h$ is a sensor compatible ZCBF.
Then any Lipschitz continuous controller
$u:\mathcal{D}\ra U$ such that $u(x)\in K(x)$ for all $x\in S$ will
render the set $\mathcal{C}$ forward invariant.
\label{cor_h_zcbf_sensing}
\end{corollary}
\begin{proof}
By assumption $u(x)\in K(x)$ for all $x\in S$.
Suppose then that $x\notin S$ so that $h(x)$ is a positive constant.
Then $K(x) = U$ since $L_fh(x) + L_gh(x)u + \alpha(h(x)) = \alpha(h(x)) > 0$
is satisfied for all $u\in U$. Hence
$u(x)\in K(x)$ for all $x\in \mathcal{D}$
so the assumptions of Theorem \ref{th_fwd_invariance} are satisfied.
\end{proof}
\begin{remark}
The difference between Theorem \ref{th_fwd_invariance} and
Corollary \ref{cor_h_zcbf_sensing} is that the condition
$u(x)\in K(x)$ only needs to be the case for $x\in S$ rather
than $x\in \mathcal{D}$ due to the extra structure on a 
sensor compatible ZCBF. This is an important distinction
because when there are sensing limitations,
it may not be possible to calculate $K(x)$ for all $x\in \mathcal{D}$.
\end{remark}

\begin{remark}
\label{rem_hstraight_not_zcbf_sensing}
For an arbitrary sensor,
neither $h_{straight}$ nor $h_{turn}$ are necessarily sensor compatible.
To see this for $h_{straight}$, let $R > 0$,
$x_1 = \begin{bmatrix} R + \epsilon & D_s & \pi & 0\end{bmatrix}^T$,
and $x_2 = \begin{bmatrix} -R + \epsilon & 0 & 0 & 0\end{bmatrix}^T$.
Then $x\notin S$ for $\epsilon > 0$ and in this case $h(x) = 0$.
However, if $x_2 = \begin{bmatrix} -R + \epsilon & -D_s & 0 & 0\end{bmatrix}^T$,
then $x\notin S$ but $h(x) = D_s$. Then $h_{straight}$ is not sensor compatible
because $h(x)$ is not constant for all $x\notin S$.
A similar calculation can be done to show $h_{turn}$ is not sensor compatible.
Hence, we cannot always apply
Corollary \ref{cor_h_zcbf_sensing} to $h_{straight}$ or $h_{turn}$ 
when there is limited range sensing.\looseness=-1
\end{remark}

Consider now some ZCBF $h$ that is not necessarily 
sensor compatible.
We now show how to create a new barrier function $\T{h}$
from $h$ so that $\T{h}$ is sensor compatible
even though this is not the case for $h$.
This allows us to be able to apply 
Corollary~\ref{cor_h_zcbf_sensing} to $\T{h}$
and keep aircraft from colliding even though they have a limited range sensor.
However, it is not always possible to create $\T{h}$ from $h$ so that $\T{h}$
is sensor compatible 
and we therefore consider two
cases. First, although $h_{turn}$ is not necessarily sensor compatible for an arbitrary sensor,
we give sufficient conditions to construct
a ZCBF to be sensor compatible.
Second, we show how to verify when it
is impossible to construct a sensor compatible ZCBF using the proposed method. We show this is
the case for $h_{straight}$.

To construct $\T{h}$, we first introduce an interpolation function
to ensure that $\T{h}$ is 
continuously differentiable, as required by the definition of a ZCBF.
Let $\xi > 0$,
$0< \beta < 1$,
and 
$\psi$ be a continuously differentiable, non-decreasing, real valued function on an open set including $[\beta\xi,\xi]$
chosen to satisfy 
\begin{equation}
\psi(\beta \xi) = \beta\xi,\quad
\psi'(\beta \xi) = 1,\quad
\psi'(\xi) = 0.
\label{eq_psi_constraints}
\end{equation}
\begin{example}
An example of $\psi$ in (\ref{eq_psi_constraints}) is 
$\psi(\eta) = c_1\eta^2 + c_2\eta + c_3$
where
$c_1 = \frac{-1}{2\xi(1 - \beta)},$ 
$c_2 = -2\xi c_1$,
and $c_3 = \beta \xi - c_1(\beta \xi)^2 - c_2 \beta \xi$.
\label{ex_quadratic_psi}
\end{example}

\begin{figure}
\centering
\centering
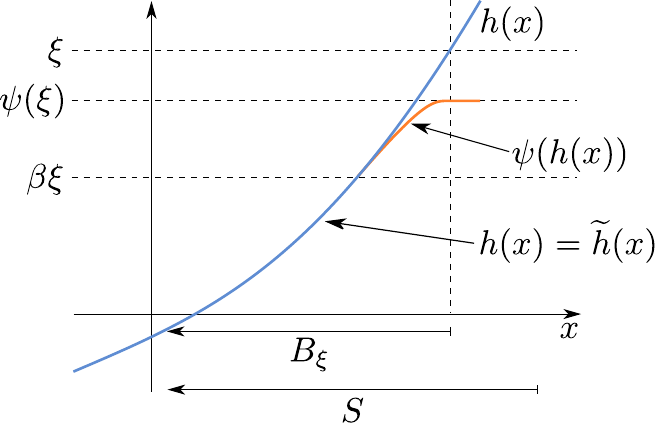
\caption{%
A graphical view of a hypothetical barrier function $h$ (blue) along with $\tilde{h}$ (orange) of a one-dimensional
system. Given $h$, the choice of $\xi$
and $\beta$ defines $\tilde{h}$ and $B_{\xi}$.
According to Theorem~\ref{th_limited_sensing_main}
when the system designer can identify a $\xi>0$ that defines
$B_\xi$ where $B_\xi\subseteq S$, $\T{h}$ is a ZCBF compatible with a sensor $s$.
\vspace*{-0.2in}
}
\label{fig_tildeh}
\end{figure}

We now define $\T{h}$ as follows
\begin{equation}
\T{h}(x) =
\begin{cases}
h(x) & h(x)\le \beta \xi \\
\psi(h(x)) & \beta \xi < h(x) < \xi \\
\psi(\xi) & \xi \le h(x)
\end{cases}
\label{eq_tilde_h_second}
\end{equation}
where we let $B_{\xi} = \set{x \in \mathcal{D} }{h(x) \le \xi}$
be a sub-level set of $h$ (see Fig. \ref{fig_tildeh})
where $\xi$ denotes the maximum value of $h$ for which
the safety constraint affects the value of $\T{h}$. 
$B_{\xi}$ is the set of states where the safety constraint
affects the value of $\T{h}$.
$\beta$ is a mixing
term for states where $\beta \xi < h(x) < \xi$ and exists to ensure the
differentiability of $\T{h}$.
\begin{remark}
With this setup, we can take the following steps to show when a ZCBF $\T{h}$
is sensor compatible.
First, the system designer chooses $\xi$ which
determines $B_\xi$. Second, the system designer determines if $B_\xi\subseteq S$.
In other words,
$\xi>0$ must be chosen with the
sensing range in mind in order to verify that $B_\xi\subseteq S$. The
second step verifies that a sensor exists so that any value of $h(x)$ such that
$h(x) < \xi$ can be calculated. Since $\T{h}$ does not require knowledge of $x$ for
$h(x) \ge \xi$, $\T{h}(x)$ can be calculated for all $x\in\mathcal{D}$.
We prove this intuition below.
Note that when multiple $\xi$ satisfy the above steps,
it may be preferable to select larger $\xi$ 
because $h(x) = \T{h}(x)$ for all $x\in\mathcal{D}$ such that $h(x) \le \beta\xi$.
\label{rem_intuition}
\end{remark}

\begin{lemma}
Assume $h$ defined in (\ref{eq_h}) is a ZCBF
where $\gamma$ is locally Lipschitz. Let $\T{h}$ be defined as in (\ref{eq_tilde_h_second}).
Then $\T{h}$ is a ZCBF on $\mathcal{D}$.
\label{lem_tildeh_is_zcbf}
\end{lemma}

\begin{proof}
Note that because of how $\psi$
is defined and because $h$ is a continuously differentiable function, that $\T{h}$ is a continuously differentiable function.
Also note that for $\beta\xi < h(x) < \xi$, $\psi(h(x)) > 0$ since $\psi$ is a non-decreasing function that is positive at $\beta \xi$.
To show that $\T{h}$ satisfies (\ref{eq_zcbf_condition}),
let $x\in \mathcal{D}$. 
We consider three cases. First, if $h(x) \le \beta \xi$
then the inequality (\ref{eq_zcbf_condition}) holds for $\T{h}(x)$ because $h(x)$
is a ZCBF. 
If $h(x) \ge \xi$ then (\ref{eq_zcbf_condition})
for $\T{h}(x)$
becomes $\alpha(\psi(\xi))\ge 0$ which is true for all $u\in U$
because $\psi(\xi) > 0$ and $\alpha$ is a class $\mathcal{K}$ function.
Finally, suppose $\beta\xi < h(x) < \xi$ and note that because $\psi$
is non-decreasing that $\pf{\psi(h(x))}{h(x)}\ge 0$. Then
\begin{IEEEeqnarray*}{l}
L_f\T{h}(x) + L_g\T{h}(x)\gamma(x) + \alpha(\T{h}(x)) \\
\qquad =
\pf{\psi(h(x))}{h(x)}\left(L_fh(x) + L_gh(x)\gamma(x)\right) + \alpha(\psi(h(x))) \\
\qquad \ge 0.
\end{IEEEeqnarray*}
The first line uses the chain rule.
The second line uses the fact that $\pf{\psi(h(x))}{h(x)}\ge 0$
and that $L_fh(x) + L_gh(x)\gamma(x) \ge 0$, as was
established in the proof of Theorem~\ref{th_h_zcbf} in \cite{squires2019composition}.
Finally, note that $\alpha(\psi(h(x))) \ge 0$
because $\psi(h(x)) > 0$ and $\alpha$ is a class $\mathcal{K}$ function.
Then $\T{h}$ is a ZCBF.
\end{proof}

\begin{theorem}
For a given sensor $S$, assume $h$ defined in (\ref{eq_h}) is a ZCBF
where $\gamma$ is locally Lipschitz and there exists a $\xi > 0$ such that
$B_{\xi} \subseteq S$.
Then $\T{h}$ defined in (\ref{eq_tilde_h_second})
is a sensor compatible ZCBF.
\label{th_limited_sensing_main}
\end{theorem}
\begin{proof}
$\T{h}$ is a ZCBF by Lemma \ref{lem_tildeh_is_zcbf}.
Suppose $x \notin S$. Then because $B_\xi \subseteq S$,
$h(x) > \xi$ so $\T{h}(x) = \psi(\xi)$. Then
$\T{h}$ is a positive constant for all $x\notin S$
and is sensor compatible.
\end{proof}

Theorem \ref{th_limited_sensing_main}
is the justification of the steps listed in Remark \ref{rem_intuition}.
Combined with Corollary \ref{cor_h_zcbf_sensing}, Theorem \ref{th_limited_sensing_main}
states how the forward invariance of a set $\mathcal{C}$ can be guaranteed
even though there is limited range sensing. However, it is predicated
on finding a $\xi > 0$ that defines a sublevel set $B_{\xi}$ for which 
$B_{\xi}\subseteq S$. We now give an example of such a case for fixed
wing collision avoidance where each aircraft is equipped with an omnidirectional
sensor with a given range $R$.\looseness=-1

\begin{example}
It was shown in Remark \ref{rem_hstraight_not_zcbf_sensing}
that $h_{turn}$ is not necessarily sensor compatible for an arbitrary sensor.
We now use Theorem \ref{th_limited_sensing_main}
to define sensing requirements so that we can create 
$\T{h}$ from $h_{turn}$ so that $\T{h}$ is sensor compatible.
For $h_{turn}$,
the trajectory defined in (\ref{eq_h_int})
is a circle for each aircraft with radius $r_1 = \sigma v / \omega$
and $r_2 = v / \omega$, respectively.
Let
$\Delta x(t) = p_{1,x}(t) - p_{2,x}(t)$,
and 
$\Delta y(t) = p_{1,y}(t) - p_{2,y}(t)$, 
so that the vehicles start at a planar distance of
$(\Delta x^2(t) + \Delta y^2(t))^{1/2}$
from each other.
Assuming the planar distance between the vehicles, $(\Delta x^2(t) + \Delta y^2(t))^{1/2}$,
is greater than $2(r_1 + r_2)$,
the closest the distance can be 
along the trajectory (\ref{eq_h_int}) is
$d_{1,2}(x(0))^{1/2} - 2r_1 - 2r_2$.
Assume each vehicle has an omnidirectional sensor with range $R$
large enough so that 
\begin{equation}
\left((R - 2r_1 - 2r_2)^2 - 4\delta\right)^{1/2} - D_s > 0.
\label{eq_R_min_sensing}
\end{equation}
Equation (\ref{eq_R_min_sensing}) implies $S$ for this example.
Having defined the sensing limitations for this problem,
we now follow the steps in Remark \ref{rem_intuition}
to show that $\T{h}$ is a sensor compatible ZCBF. First,
we choose $\xi$ so that we can prove $B_\xi \subseteq S$, namely
\begin{equation}
\left((R - 2r_1 - 2r_2)^2 - 4\delta\right)^{1/2} - D_s = \xi > 0.
\label{eq_xi_min_sensing}
\end{equation}
Second, we show that $B_{\xi} \subseteq S$.
Suppose $x(t)\notin S$ so that $d_{1,2}(x(t)) > R^2$.
Then because the trajectories of each vehicle
is a circle in (\ref{eq_h_int}),
\begin{IEEEeqnarray*}{rCl}
h(x(t))
&=& \INF \rho(t + \tau) \\
&\ge& ((d_{1,2}(x(0))^{1/2} - 2r_1 - 2r_2)^2 - 4\delta)^{1/2} - D_s \\
&>& ((R - 2r_1 - 2r_2)^2 - 4\delta)^{1/2} - D_s = \xi > 0.
\end{IEEEeqnarray*}
Then $x(t)\notin B_{\xi}$. Then $B_{\xi}\subseteq S$.
In other words, given a sensor of radius $R$,
we can choose $\xi$ according to (\ref{eq_xi_min_sensing})
and use $\T{h}$ defined in (\ref{eq_tilde_h_second}) to
ensure safe operations between two UAVs.\looseness=-1
\label{ex_fixed_wing_lim_range_safety}
\end{example}

While Example~\ref{ex_fixed_wing_lim_range_safety} showed how to
use $h_{turn}$ with Theorem~\ref{th_limited_sensing_main}
in order to ensure UAV collision avoidance in spite of limited range sensing,
the same cannot be done for $h_{straight}$.
\begin{corollary}
Assume $h$ defined in (\ref{eq_h}) is a ZCBF
where $\gamma$ is locally Lipschitz.
Suppose there exists an $x\in\mathcal{D}$ such that $h(x) < 0$ and $x\notin S$.
Then for all $\xi > 0$, $B_\xi \not\subseteq  S$.
\label{cor_cant_use_main_theorem}
\end{corollary}
\begin{proof}
Note that for the given $x$, $x\in B_\xi$ for any $\xi > 0$ since $h(x) < 0$.
Then $x\in B_\xi$ but $x\not\in S$.
\end{proof}
\begin{remark}
We now use
Corollary \ref{cor_cant_use_main_theorem} to show
$h_{straight}$ cannot be used with Theorem \ref{th_limited_sensing_main}
to guarantee safety. Let
$x_1(0) = \begin{bmatrix} -R / 2 - \epsilon & 0 & 0 & 0\end{bmatrix}^T$,
$x_2(0) = \begin{bmatrix} R / 2 + \epsilon & 0 & \pi & 0\end{bmatrix}^T$
the sensing radius be $R$, and $\epsilon > 0$.
Then $h(x) = -D_s$
and $x\not\in S$ since the vehicles are further than $R$ apart.
\end{remark}

%% file: img/svg/tildeh.pdf_tex
%% Accompanies image file 'img/svg/tildeh.pdf' (pdf, eps, ps)
%%   \input{<filename>img/svg/.pdf_tex}
%%   \includegraphics{<filename>img/svg/.pdf}
%%   \input{<filename>img/svg/.pdf_tex}
%%   \includegraphics[width=<desired width>]{<filename>img/svg/.pdf}
%%   \import{<path to file>}{<filename>img/svg/.pdf_tex}
\begingroup%
  \makeatletter%
  \providecommand\color[2][]{%
    \errmessage{(Inkscape) Color is used for the text in Inkscape, but the package 'color.sty' is not loaded}%
    \renewcommand\color[2][]{}%
  }%
  \providecommand\transparent[1]{%
    \errmessage{(Inkscape) Transparency is used (non-zero) for the text in Inkscape, but the package 'transparent.sty' is not loaded}%
    \renewcommand\transparent[1]{}%
  }%
  \providecommand\rotatebox[2]{#2}%
  \newcommand*\fsize{\dimexpr\f@size pt\relax}%
  \newcommand*\lineheight[1]{\fontsize{\fsize}{#1\fsize}\selectfont}%
  \ifx\svgwidth\undefined%
    \setlength{\unitlength}{188.4544253bp}%
    \ifx\svgscale\undefined%
      \relax%
    \else%
      \setlength{\unitlength}{\unitlength * \real{\svgscale}}%
    \fi%
  \else%
    \setlength{\unitlength}{\svgwidth}%
  \fi%
  \global\let\svgwidth\undefined%
  \global\let\svgscale\undefined%
  \makeatother%
  \begin{picture}(1,0.64597748)%
    \lineheight{1}%
    \setlength\tabcolsep{0pt}%
    \put(0,0){\includegraphics[width=\unitlength,page=1]{img/svg/tildeh.pdf}}%
  \end{picture}%
\endgroup%

%% file: alpha.tex
\section{AN INTERPRETATION OF $\T{h}$ AS A MORE PERMISSIVE ZCBF THAN $h$}
\label{sec_alpha}

Consider a ZCBF $h$ that is not necessarily sensor compatible
but for which it is possible to construct $\T{h}$
so that $\T{h}$ is a sensor compatible ZCBF.
In this
section we characterize how $\T{h}$ is more permissive than $h$.
For notational convenience
we denote $\nabla_h\psi(h(x)) = \pf{\psi(h(x))}{h(x)}$ and assume the following.

\begin{assumption}
Assume on $(\beta\xi,\xi)$,
$\alpha$ is continuously differentiable. Further,
assume the derivative of $\alpha$ is non-increasing on $(\beta\xi,\xi)$.
\label{as_alpha}
\end{assumption}
\begin{remark}
The assumptions on $\alpha$ can be satisfied for any $\alpha$
that is linear on the region $(\beta\xi,\xi)$.
\end{remark}
\begin{assumption}
Assume the domain of $\psi$ is extended by letting $\psi$ be the identity function for inputs $h(x) < \beta\xi$.
Further, assume
the first derivative of $\psi$ is strictly positive but non-increasing on $(\beta\xi,\xi)$
and the second derivative of $\psi$ is negative on $(\beta\xi,\xi)$.
\label{as_psi}
\end{assumption}
\begin{remark}
Note that because the first derivative of $\psi$ is strictly positive for $h(x) < \xi$,
$(\nabla_h\psi(h(x)))^{-1}$ is well defined on $h(x)< \xi$.
\label{rem_psi_der_defined}
\end{remark}
\begin{remark}
The $\psi$ discussed in Example \ref{ex_quadratic_psi}
satisfies Assumption \ref{as_psi}
by letting $\psi(h(x)) = h(x)$ for $h(x) \le \beta \xi$.
\end{remark}

We begin by expanding the barrier condition (\ref{def_zcbf}) for $\T{h}$
in terms of $h$.  This will
then be used to show the conditions such that $K(x)\subseteq \T{K}(x)$
where $\T{K}(x)$ is the admissible control space of $\T{h}$ at $x$.
Under Assumption~\ref{as_psi} and noting Remark \ref{rem_psi_der_defined}, for
$h(x)< \xi$ we have for $x\in\mathcal{D}$ that
\begin{IEEEeqnarray*}{l}
L_f\T{h}(x) + L_g\T{h}(x)u + \alpha(\T{h}(x)) \\
\quad = \nabla_h \psi(h(x)) \left(L_fh(x) + L_gh(x)u\right) + \alpha(\psi(h(x))) \\
\quad = \nabla_h \psi(h(x)) \Big(L_fh(x) + L_gh(x)u + \\
\qquad \qquad\qquad\qquad   (\nabla_h\psi(h(x)))^{-1}\alpha(\psi(h(x)))\Big).
\end{IEEEeqnarray*}
Then because $\nabla_h\psi(h(x)) > 0$ and letting 
\begin{equation}
\alpha_2(h(x)) = (\nabla_h\psi(h(x)))^{-1}\alpha(\psi(h(x))),
\label{eq_alpha2}
\end{equation}
we have, letting $\sign$ be the sign of the expression,
\begin{IEEEeqnarray*}{l}
\sign(L_f\T{h}(x) + L_g\T{h}(x)u + \alpha(\T{h}(x))) \IEEEyesnumber\label{eq_alpha}\\
\quad = 
\sign(L_fh(x) + L_gh(x)u + \alpha_2(h(x))).
\end{IEEEeqnarray*}

\begin{lemma}
Suppose $h$ is a ZCBF, let $\T{h}$ be as defined in (\ref{eq_tilde_h_second}),
let Assumptions \ref{as_alpha} and \ref{as_psi} hold,
and let $\alpha_2$ be as defined in (\ref{eq_alpha2}).
If $h(x) < \xi$ then $\alpha_2(h(x)) \ge \alpha(h(x))$.
\label{lem_alpha}
\end{lemma}
\begin{proof}
Suppose $h(x) \le \beta\xi$. Then because $\psi(h(x)) = h(x)$,
$\alpha_2(h(x)) = \alpha(h(x))$.
Suppose now that $\beta\xi < h(x) < \xi$. We
prove $\alpha_2(h(x)) \ge \alpha(h(x))$
with the comparison lemma \cite{hassan2002nonlinear}.\looseness=-1

It has already been shown that 
$\alpha_2(h(x)) = \alpha(h(x))$ at $h(x) = \beta\xi$.
We now show that
$\nabla_h \alpha_2(h(x)) \ge \nabla_h \alpha(h(x))$ for $h(x)\in (\beta\xi,\xi)$.
By the chain rule,
\begin{IEEEeqnarray*}{l}
\nabla_h \alpha_2(h(x)) \\
\quad =
-\left(\frac{1}{\nabla_h\psi(h(x))}\right)^2\nabla^2_h \psi(h(x)) \alpha(\psi(h(x)) \\
\qquad + \left(\frac{1}{\nabla_h\psi(h(x))}\right)\nabla_\psi \alpha(\psi(h(x)))\nabla_h \psi(h(x)) \\
\quad \ge \nabla_\psi \alpha(\psi(h(x))).
\end{IEEEeqnarray*}
The inequality holds because the second derivative of $\psi$
is negative and $\alpha(\psi(h(x)) \ge 0$ for $h(x)\in(\beta\xi,\xi)$.
We must now show that $\nabla_\psi \alpha(\psi(h(x))) \ge \nabla_h \alpha(h(x))$
to conclude that $\alpha_2(h(x)) \ge \alpha(h(x))$
for $h(x)\in(\beta\xi,\xi)$.

Because $\psi(h(x)) = h(x)$ for $h(x) = \beta\xi$, the first derivative of $\psi$ is 1 at $h(x) = \beta\xi$
and the first derivative is non-increasing on $h(x) \in (\beta\xi,\xi)$,
so
$\psi(h(x))\le h(x)$ for $h(x) \in (\beta\xi,\xi)$.
Then because the derivative of $\alpha$ is non-increasing
for $h(x)\in(\beta\xi,\xi)$,
$\nabla_\psi \alpha(\psi(h(x)) \ge \nabla_h \alpha(h(x))$.

\end{proof}
\begin{remark}
\vspace*{-0.1in}
Note that $\alpha_2$ is a class $\mathcal{K}$ function.
By definition, $\alpha_2(0) = 0$ and
is strictly increasing on $(0,\beta\xi)$. To see that
$\alpha_2$ is strictly increasing on $(\beta\xi,\xi)$,
note that it has already been proven that
$\nabla_h \alpha_2(h(x)) \ge \nabla_\psi\alpha(\psi(h(x))) \ge \nabla_h \alpha(h(x))$.
Further $\nabla_h\alpha(h(x)) > 0$ since
$\alpha$ is a class $\mathcal{K}$ function.
Then $\nabla_h \alpha_2(h(x)) > 0$.
\end{remark}

\begin{theorem}
Suppose $h$ is a ZCBF, assume $\T{h}$ as defined in (\ref{eq_tilde_h_second})
is sensor compatible,
and let Assumptions \ref{as_alpha} and \ref{as_psi} hold.
Then $K(x) \subseteq \T{K}(x)$ for all $x\in\mathcal{D}$.
\label{th_kx_in_tkx}
\end{theorem}
\begin{proof}
Suppose $x$ is such that $h(x) < \xi$ and $u\in K(x)$.
Then $L_fh(x) + L_gh(x)u + \alpha(h(x)) \ge 0$.
Then since $\alpha_2(h(x)) \ge \alpha(h(x))$ from Lemma \ref{lem_alpha}
we have $L_fh(x) + L_fh(x)u + \alpha_2(h(x)) \ge 0$.
Then from (\ref{eq_alpha}), $L_f\T{h}(x) + L_g\T{h}(x)u + \alpha(\T{h}(x)) \ge 0$.
Then $u\in\T{K}(x)$.

Suppose $x$ is such that $h(x) \ge \xi$
and $u\in K(x)$. Then because $u\in U$, $u\in\T{K}(x)$ since $\T{K}(x) = U$.
\end{proof}
\begin{remark}
In particular, Theorem \ref{th_kx_in_tkx} gives the conditions under which any $u$
satisfying the QP (\ref{eq_general_qp}) using $h$ will be satisfied
in a QP (\ref{eq_general_qp}) when using $\T{h}$.
\end{remark}

%% file: sim.tex
\section{SIMULATION EXPERIMENTS}
\label{sec_simulation}
In this section we conduct a simulation experiment
with SCRIMMAGE \cite{demarco2018}.
We consider two vehicles with initial states of
$\begin{bmatrix} -200 & 0 & 0 & 0\end{bmatrix}^T$
and
$\begin{bmatrix} 200 & 0 & \pi & 0\end{bmatrix}^T$
with goal positions of
$\begin{bmatrix} 200 & 0 & 0\end{bmatrix}^T$
and
$\begin{bmatrix} -200 & 0 & 0 \end{bmatrix}^T$,
respectively.
We use
$h_{turn}$, letting
$v = 0.9 v_{min} + 0.1v_{max}$ and $\omega = 0.9 \omega_{max}$ in (\ref{eq_gamma_turn})
where $v_{min} = 15$ meters/second, $v_{max} = 25$ meters/second, $\delta = 0.01$ meters$^2$,
and $\omega_{max} = 13$ degrees/second.
The choice of $\omega_{max}$
results from assuming a $30$ degree max bank angle while traveling at $v_{max}$
and using a constant rate turn formula, as in \cite{clancy1975aerodynamics}.

\begin{figure}[t]
\begin{center}
\input{img/sim/sensing.pgf}
\caption{The minimum vehicle distance vs sensing range.
The green dashed line is the minimum sensing range
using (\ref{eq_R_min_sensing}). Note that for all sensing ranges
above the minimum sensing range, the vehicles are able to
maintain safe distances.
\vspace*{-0.3in}}
\label{fig_sensing_distance}
\end{center}
\end{figure}
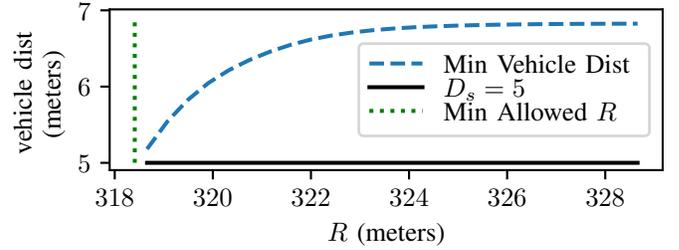

In the first experiment, we examine the effect of sensing range
on the resulting closest distance the vehicles experience during the
simulation.
In Example \ref{ex_fixed_wing_lim_range_safety} we showed
how to apply the steps described in Remark \ref{rem_intuition}
by choosing $\xi$ so that $B_\xi\subseteq S$
to conclude that 
that $\T{h}$ is sensor compatible. The conclusion
required that the sensing range was above a threshold in (\ref{eq_R_min_sensing}).
Using
the parameters of this experiment, equation (\ref{eq_R_min_sensing}) implies $R > 318.4$.
Because the inequality is strict, we start the experiment
with $R = 319$.
As shown in Fig.~\ref{fig_sensing_distance}, provided
the sensing range is above the threshold calculated in (\ref{eq_R_min_sensing}),
the vehicles are able to maintain safe distances throughout the simulation.
Further, as the sensing range approaches the limit predicted by (\ref{eq_R_min_sensing}),
the minimum distance between the vehicles approaches
$D_s$.

In the second experiment we repeat the experiment of
\cite{squires2019composition} where 20 vehicles are applying a barrier function and are positioned around a circle
with a nominal controller that cause the vehicles to arrive at the origin at
the same time.  We note that satisfying multiple constraints simultaneously
with barrier certificates has been previously addressed
\cite{xu2017correctness,wang2018permissive,xu2018constrained,squires2019composition}
and we use the method discussed in \cite{squires2019composition} for this
experiment.  The difference in this experiment from
\cite{squires2019composition} is that we include a limited sensing range of
$350$ for each vehicle and start the vehicles $1250$ feet from the origin so
they start the scenario without being able to sense each other. A
video simulation is shown in \cite{squires2021sensingvideo} and the vehicles are able to maintain safe
distances throughout the simulation.

% \begin{figure}
% \centering
% \begin{subfigure}{\columnwidth}
% \centering
% \input{img/sim/20_veh1_actuator0.pgf}
% \caption{}
% \end{subfigure}
% 
% \begin{subfigure}{\columnwidth}
% \centering
% \input{img/sim/20_veh1_actuator1.pgf}
% \caption{}
% \end{subfigure}
% 
% \begin{subfigure}{\columnwidth}
% \centering
% \input{img/sim/20_veh1_actuator2.pgf}
% \caption{}
% \end{subfigure}
% 
% \begin{subfigure}{\columnwidth}
% \centering
% \input{img/sim/20_veh_min_d.pgf}
% \caption{}
% \end{subfigure}
% 
% \begin{subfigure}{\columnwidth}
% \centering
% \input{img/sim/20_veh1_path.pgf}
% \caption{}
% \end{subfigure}
% 
% \caption[Control outputs for the 20 vehicle case]{
% Outputs for the scenario with 20 fixed-wing vehicles.
% The blue dashed and orange
% correspond to the case where the vehicles have unlimited sensing
% and a sensing radius of 350 meters, respectively.
% Actuator outputs are shown in to be within the constrains in (a), (b) and (c) for vehicle 1. In (d), it is shown
% that all vehicles maintain safe distances from each other, verifying 
% that the algorithm has kept the vehicles safe. In (e), the path
% of vehicle 1 is shown. Note that the blue dashed and orange lines
% are almost on top of each other, indicating very similar performance
% when the vehicles have limited sensing.}
% 
% \label{fig_constraints_satisfied}
% \end{figure}

%% file: img/sim/sensing.pgf
%% Creator: Matplotlib, PGF backend
%%
%% To include the figure in your LaTeX document, write
%%   \input{<filename>.pgf}
%%
%% Make sure the required packages are loaded in your preamble
%%   \usepackage{pgf}
%%
%% and, on pdftex
%%   \usepackage[utf8]{inputenc}\DeclareUnicodeCharacter{2212}{-}
%%
%% or, on luatex and xetex
%%   \usepackage{unicode-math}
%%
%% Figures using additional raster images can only be included by \input if
%% they are in the same directory as the main LaTeX file. For loading figures
%% from other directories you can use the `import` package
%% and then include the figures with
%%   \import{<path to file>}{<filename>.pgf}
%%
%% Matplotlib used the following preamble
%%   \usepackage[utf8x]{inputenc}
%%   \usepackage[T1]{fontenc}
%%   \newcommand{\vect}[1]{#1}
%%   \usepackage{fontspec}
%%
\begingroup%
\makeatletter%
\begin{pgfpicture}%
\pgfpathrectangle{\pgfpointorigin}{\pgfqpoint{3.400000in}{1.300000in}}%
\pgfusepath{use as bounding box, clip}%
\begin{pgfscope}%
\pgfsetbuttcap%
\pgfsetmiterjoin%
\definecolor{currentfill}{rgb}{1.000000,1.000000,1.000000}%
\pgfsetfillcolor{currentfill}%
\pgfsetlinewidth{0.000000pt}%
\definecolor{currentstroke}{rgb}{1.000000,1.000000,1.000000}%
\pgfsetstrokecolor{currentstroke}%
\pgfsetdash{}{0pt}%
\pgfpathmoveto{\pgfqpoint{0.000000in}{0.000000in}}%
\pgfpathlineto{\pgfqpoint{3.400000in}{0.000000in}}%
\pgfpathlineto{\pgfqpoint{3.400000in}{1.300000in}}%
\pgfpathlineto{\pgfqpoint{0.000000in}{1.300000in}}%
\pgfpathclose%
\pgfusepath{fill}%
\end{pgfscope}%
\begin{pgfscope}%
\pgfsetbuttcap%
\pgfsetmiterjoin%
\definecolor{currentfill}{rgb}{1.000000,1.000000,1.000000}%
\pgfsetfillcolor{currentfill}%
\pgfsetlinewidth{0.000000pt}%
\definecolor{currentstroke}{rgb}{0.000000,0.000000,0.000000}%
\pgfsetstrokecolor{currentstroke}%
\pgfsetstrokeopacity{0.000000}%
\pgfsetdash{}{0pt}%
\pgfpathmoveto{\pgfqpoint{0.505278in}{0.415000in}}%
\pgfpathlineto{\pgfqpoint{3.400000in}{0.415000in}}%
\pgfpathlineto{\pgfqpoint{3.400000in}{1.256836in}}%
\pgfpathlineto{\pgfqpoint{0.505278in}{1.256836in}}%
\pgfpathclose%
\pgfusepath{fill}%
\end{pgfscope}%
\begin{pgfscope}%
\pgfsetbuttcap%
\pgfsetroundjoin%
\definecolor{currentfill}{rgb}{0.000000,0.000000,0.000000}%
\pgfsetfillcolor{currentfill}%
\pgfsetlinewidth{0.803000pt}%
\definecolor{currentstroke}{rgb}{0.000000,0.000000,0.000000}%
\pgfsetstrokecolor{currentstroke}%
\pgfsetdash{}{0pt}%
\pgfsys@defobject{currentmarker}{\pgfqpoint{0.000000in}{-0.048611in}}{\pgfqpoint{0.000000in}{0.000000in}}{%
\pgfpathmoveto{\pgfqpoint{0.000000in}{0.000000in}}%
\pgfpathlineto{\pgfqpoint{0.000000in}{-0.048611in}}%
\pgfusepath{stroke,fill}%
}%
\begin{pgfscope}%
\pgfsys@transformshift{0.531899in}{0.415000in}%
\pgfsys@useobject{currentmarker}{}%
\end{pgfscope}%
\end{pgfscope}%
\begin{pgfscope}%
\definecolor{textcolor}{rgb}{0.000000,0.000000,0.000000}%
\pgfsetstrokecolor{textcolor}%
\pgfsetfillcolor{textcolor}%
\pgftext[x=0.531899in,y=0.317777in,,top]{\color{textcolor}\rmfamily\fontsize{10.000000}{12.000000}\selectfont \(\displaystyle {318}\)}%
\end{pgfscope}%
\begin{pgfscope}%
\pgfsetbuttcap%
\pgfsetroundjoin%
\definecolor{currentfill}{rgb}{0.000000,0.000000,0.000000}%
\pgfsetfillcolor{currentfill}%
\pgfsetlinewidth{0.803000pt}%
\definecolor{currentstroke}{rgb}{0.000000,0.000000,0.000000}%
\pgfsetstrokecolor{currentstroke}%
\pgfsetdash{}{0pt}%
\pgfsys@defobject{currentmarker}{\pgfqpoint{0.000000in}{-0.048611in}}{\pgfqpoint{0.000000in}{0.000000in}}{%
\pgfpathmoveto{\pgfqpoint{0.000000in}{0.000000in}}%
\pgfpathlineto{\pgfqpoint{0.000000in}{-0.048611in}}%
\pgfusepath{stroke,fill}%
}%
\begin{pgfscope}%
\pgfsys@transformshift{1.045375in}{0.415000in}%
\pgfsys@useobject{currentmarker}{}%
\end{pgfscope}%
\end{pgfscope}%
\begin{pgfscope}%
\definecolor{textcolor}{rgb}{0.000000,0.000000,0.000000}%
\pgfsetstrokecolor{textcolor}%
\pgfsetfillcolor{textcolor}%
\pgftext[x=1.045375in,y=0.317777in,,top]{\color{textcolor}\rmfamily\fontsize{10.000000}{12.000000}\selectfont \(\displaystyle {320}\)}%
\end{pgfscope}%
\begin{pgfscope}%
\pgfsetbuttcap%
\pgfsetroundjoin%
\definecolor{currentfill}{rgb}{0.000000,0.000000,0.000000}%
\pgfsetfillcolor{currentfill}%
\pgfsetlinewidth{0.803000pt}%
\definecolor{currentstroke}{rgb}{0.000000,0.000000,0.000000}%
\pgfsetstrokecolor{currentstroke}%
\pgfsetdash{}{0pt}%
\pgfsys@defobject{currentmarker}{\pgfqpoint{0.000000in}{-0.048611in}}{\pgfqpoint{0.000000in}{0.000000in}}{%
\pgfpathmoveto{\pgfqpoint{0.000000in}{0.000000in}}%
\pgfpathlineto{\pgfqpoint{0.000000in}{-0.048611in}}%
\pgfusepath{stroke,fill}%
}%
\begin{pgfscope}%
\pgfsys@transformshift{1.558851in}{0.415000in}%
\pgfsys@useobject{currentmarker}{}%
\end{pgfscope}%
\end{pgfscope}%
\begin{pgfscope}%
\definecolor{textcolor}{rgb}{0.000000,0.000000,0.000000}%
\pgfsetstrokecolor{textcolor}%
\pgfsetfillcolor{textcolor}%
\pgftext[x=1.558851in,y=0.317777in,,top]{\color{textcolor}\rmfamily\fontsize{10.000000}{12.000000}\selectfont \(\displaystyle {322}\)}%
\end{pgfscope}%
\begin{pgfscope}%
\pgfsetbuttcap%
\pgfsetroundjoin%
\definecolor{currentfill}{rgb}{0.000000,0.000000,0.000000}%
\pgfsetfillcolor{currentfill}%
\pgfsetlinewidth{0.803000pt}%
\definecolor{currentstroke}{rgb}{0.000000,0.000000,0.000000}%
\pgfsetstrokecolor{currentstroke}%
\pgfsetdash{}{0pt}%
\pgfsys@defobject{currentmarker}{\pgfqpoint{0.000000in}{-0.048611in}}{\pgfqpoint{0.000000in}{0.000000in}}{%
\pgfpathmoveto{\pgfqpoint{0.000000in}{0.000000in}}%
\pgfpathlineto{\pgfqpoint{0.000000in}{-0.048611in}}%
\pgfusepath{stroke,fill}%
}%
\begin{pgfscope}%
\pgfsys@transformshift{2.072328in}{0.415000in}%
\pgfsys@useobject{currentmarker}{}%
\end{pgfscope}%
\end{pgfscope}%
\begin{pgfscope}%
\definecolor{textcolor}{rgb}{0.000000,0.000000,0.000000}%
\pgfsetstrokecolor{textcolor}%
\pgfsetfillcolor{textcolor}%
\pgftext[x=2.072328in,y=0.317777in,,top]{\color{textcolor}\rmfamily\fontsize{10.000000}{12.000000}\selectfont \(\displaystyle {324}\)}%
\end{pgfscope}%
\begin{pgfscope}%
\pgfsetbuttcap%
\pgfsetroundjoin%
\definecolor{currentfill}{rgb}{0.000000,0.000000,0.000000}%
\pgfsetfillcolor{currentfill}%
\pgfsetlinewidth{0.803000pt}%
\definecolor{currentstroke}{rgb}{0.000000,0.000000,0.000000}%
\pgfsetstrokecolor{currentstroke}%
\pgfsetdash{}{0pt}%
\pgfsys@defobject{currentmarker}{\pgfqpoint{0.000000in}{-0.048611in}}{\pgfqpoint{0.000000in}{0.000000in}}{%
\pgfpathmoveto{\pgfqpoint{0.000000in}{0.000000in}}%
\pgfpathlineto{\pgfqpoint{0.000000in}{-0.048611in}}%
\pgfusepath{stroke,fill}%
}%
\begin{pgfscope}%
\pgfsys@transformshift{2.585804in}{0.415000in}%
\pgfsys@useobject{currentmarker}{}%
\end{pgfscope}%
\end{pgfscope}%
\begin{pgfscope}%
\definecolor{textcolor}{rgb}{0.000000,0.000000,0.000000}%
\pgfsetstrokecolor{textcolor}%
\pgfsetfillcolor{textcolor}%
\pgftext[x=2.585804in,y=0.317777in,,top]{\color{textcolor}\rmfamily\fontsize{10.000000}{12.000000}\selectfont \(\displaystyle {326}\)}%
\end{pgfscope}%
\begin{pgfscope}%
\pgfsetbuttcap%
\pgfsetroundjoin%
\definecolor{currentfill}{rgb}{0.000000,0.000000,0.000000}%
\pgfsetfillcolor{currentfill}%
\pgfsetlinewidth{0.803000pt}%
\definecolor{currentstroke}{rgb}{0.000000,0.000000,0.000000}%
\pgfsetstrokecolor{currentstroke}%
\pgfsetdash{}{0pt}%
\pgfsys@defobject{currentmarker}{\pgfqpoint{0.000000in}{-0.048611in}}{\pgfqpoint{0.000000in}{0.000000in}}{%
\pgfpathmoveto{\pgfqpoint{0.000000in}{0.000000in}}%
\pgfpathlineto{\pgfqpoint{0.000000in}{-0.048611in}}%
\pgfusepath{stroke,fill}%
}%
\begin{pgfscope}%
\pgfsys@transformshift{3.099280in}{0.415000in}%
\pgfsys@useobject{currentmarker}{}%
\end{pgfscope}%
\end{pgfscope}%
\begin{pgfscope}%
\definecolor{textcolor}{rgb}{0.000000,0.000000,0.000000}%
\pgfsetstrokecolor{textcolor}%
\pgfsetfillcolor{textcolor}%
\pgftext[x=3.099280in,y=0.317777in,,top]{\color{textcolor}\rmfamily\fontsize{10.000000}{12.000000}\selectfont \(\displaystyle {328}\)}%
\end{pgfscope}%
\begin{pgfscope}%
\definecolor{textcolor}{rgb}{0.000000,0.000000,0.000000}%
\pgfsetstrokecolor{textcolor}%
\pgfsetfillcolor{textcolor}%
\pgftext[x=1.952639in,y=0.138889in,,top]{\color{textcolor}\rmfamily\fontsize{10.000000}{12.000000}\selectfont \(\displaystyle R\) (meters)}%
\end{pgfscope}%
\begin{pgfscope}%
\pgfsetbuttcap%
\pgfsetroundjoin%
\definecolor{currentfill}{rgb}{0.000000,0.000000,0.000000}%
\pgfsetfillcolor{currentfill}%
\pgfsetlinewidth{0.803000pt}%
\definecolor{currentstroke}{rgb}{0.000000,0.000000,0.000000}%
\pgfsetstrokecolor{currentstroke}%
\pgfsetdash{}{0pt}%
\pgfsys@defobject{currentmarker}{\pgfqpoint{-0.048611in}{0.000000in}}{\pgfqpoint{-0.000000in}{0.000000in}}{%
\pgfpathmoveto{\pgfqpoint{-0.000000in}{0.000000in}}%
\pgfpathlineto{\pgfqpoint{-0.048611in}{0.000000in}}%
\pgfusepath{stroke,fill}%
}%
\begin{pgfscope}%
\pgfsys@transformshift{0.505278in}{0.453265in}%
\pgfsys@useobject{currentmarker}{}%
\end{pgfscope}%
\end{pgfscope}%
\begin{pgfscope}%
\definecolor{textcolor}{rgb}{0.000000,0.000000,0.000000}%
\pgfsetstrokecolor{textcolor}%
\pgfsetfillcolor{textcolor}%
\pgftext[x=0.338611in, y=0.405070in, left, base]{\color{textcolor}\rmfamily\fontsize{10.000000}{12.000000}\selectfont \(\displaystyle {5}\)}%
\end{pgfscope}%
\begin{pgfscope}%
\pgfsetbuttcap%
\pgfsetroundjoin%
\definecolor{currentfill}{rgb}{0.000000,0.000000,0.000000}%
\pgfsetfillcolor{currentfill}%
\pgfsetlinewidth{0.803000pt}%
\definecolor{currentstroke}{rgb}{0.000000,0.000000,0.000000}%
\pgfsetstrokecolor{currentstroke}%
\pgfsetdash{}{0pt}%
\pgfsys@defobject{currentmarker}{\pgfqpoint{-0.048611in}{0.000000in}}{\pgfqpoint{-0.000000in}{0.000000in}}{%
\pgfpathmoveto{\pgfqpoint{-0.000000in}{0.000000in}}%
\pgfpathlineto{\pgfqpoint{-0.048611in}{0.000000in}}%
\pgfusepath{stroke,fill}%
}%
\begin{pgfscope}%
\pgfsys@transformshift{0.505278in}{0.852538in}%
\pgfsys@useobject{currentmarker}{}%
\end{pgfscope}%
\end{pgfscope}%
\begin{pgfscope}%
\definecolor{textcolor}{rgb}{0.000000,0.000000,0.000000}%
\pgfsetstrokecolor{textcolor}%
\pgfsetfillcolor{textcolor}%
\pgftext[x=0.338611in, y=0.804344in, left, base]{\color{textcolor}\rmfamily\fontsize{10.000000}{12.000000}\selectfont \(\displaystyle {6}\)}%
\end{pgfscope}%
\begin{pgfscope}%
\pgfsetbuttcap%
\pgfsetroundjoin%
\definecolor{currentfill}{rgb}{0.000000,0.000000,0.000000}%
\pgfsetfillcolor{currentfill}%
\pgfsetlinewidth{0.803000pt}%
\definecolor{currentstroke}{rgb}{0.000000,0.000000,0.000000}%
\pgfsetstrokecolor{currentstroke}%
\pgfsetdash{}{0pt}%
\pgfsys@defobject{currentmarker}{\pgfqpoint{-0.048611in}{0.000000in}}{\pgfqpoint{-0.000000in}{0.000000in}}{%
\pgfpathmoveto{\pgfqpoint{-0.000000in}{0.000000in}}%
\pgfpathlineto{\pgfqpoint{-0.048611in}{0.000000in}}%
\pgfusepath{stroke,fill}%
}%
\begin{pgfscope}%
\pgfsys@transformshift{0.505278in}{1.251811in}%
\pgfsys@useobject{currentmarker}{}%
\end{pgfscope}%
\end{pgfscope}%
\begin{pgfscope}%
\definecolor{textcolor}{rgb}{0.000000,0.000000,0.000000}%
\pgfsetstrokecolor{textcolor}%
\pgfsetfillcolor{textcolor}%
\pgftext[x=0.338611in, y=1.203617in, left, base]{\color{textcolor}\rmfamily\fontsize{10.000000}{12.000000}\selectfont \(\displaystyle {7}\)}%
\end{pgfscope}%
\begin{pgfscope}%
\definecolor{textcolor}{rgb}{0.000000,0.000000,0.000000}%
\pgfsetstrokecolor{textcolor}%
\pgfsetfillcolor{textcolor}%
\pgftext[x=0.096389in, y=0.496057in, left, base,rotate=90.000000]{\color{textcolor}\rmfamily\fontsize{10.000000}{12.000000}\selectfont vehicle dist}%
\end{pgfscope}%
\begin{pgfscope}%
\definecolor{textcolor}{rgb}{0.000000,0.000000,0.000000}%
\pgfsetstrokecolor{textcolor}%
\pgfsetfillcolor{textcolor}%
\pgftext[x=0.248333in, y=0.580779in, left, base,rotate=90.000000]{\color{textcolor}\rmfamily\fontsize{10.000000}{12.000000}\selectfont (meters)}%
\end{pgfscope}%
\begin{pgfscope}%
\pgfpathrectangle{\pgfqpoint{0.505278in}{0.415000in}}{\pgfqpoint{2.894722in}{0.841837in}}%
\pgfusepath{clip}%
\pgfsetbuttcap%
\pgfsetroundjoin%
\pgfsetlinewidth{1.505625pt}%
\definecolor{currentstroke}{rgb}{0.121569,0.466667,0.705882}%
\pgfsetstrokecolor{currentstroke}%
\pgfsetdash{{5.550000pt}{2.400000pt}}{0.000000pt}%
\pgfpathmoveto{\pgfqpoint{0.701040in}{0.525475in}}%
\pgfpathlineto{\pgfqpoint{0.808015in}{0.670339in}}%
\pgfpathlineto{\pgfqpoint{0.914989in}{0.783400in}}%
\pgfpathlineto{\pgfqpoint{1.021963in}{0.871272in}}%
\pgfpathlineto{\pgfqpoint{1.128937in}{0.938348in}}%
\pgfpathlineto{\pgfqpoint{1.235912in}{0.991470in}}%
\pgfpathlineto{\pgfqpoint{1.342886in}{1.034902in}}%
\pgfpathlineto{\pgfqpoint{1.449860in}{1.070617in}}%
\pgfpathlineto{\pgfqpoint{1.556834in}{1.098707in}}%
\pgfpathlineto{\pgfqpoint{1.663808in}{1.119662in}}%
\pgfpathlineto{\pgfqpoint{1.770783in}{1.135452in}}%
\pgfpathlineto{\pgfqpoint{1.877757in}{1.147384in}}%
\pgfpathlineto{\pgfqpoint{1.984731in}{1.156228in}}%
\pgfpathlineto{\pgfqpoint{2.091705in}{1.162954in}}%
\pgfpathlineto{\pgfqpoint{2.198680in}{1.167756in}}%
\pgfpathlineto{\pgfqpoint{2.305654in}{1.171430in}}%
\pgfpathlineto{\pgfqpoint{2.412628in}{1.174256in}}%
\pgfpathlineto{\pgfqpoint{2.519602in}{1.176432in}}%
\pgfpathlineto{\pgfqpoint{2.626576in}{1.178102in}}%
\pgfpathlineto{\pgfqpoint{2.733551in}{1.179377in}}%
\pgfpathlineto{\pgfqpoint{2.840525in}{1.180342in}}%
\pgfpathlineto{\pgfqpoint{2.947499in}{1.181065in}}%
\pgfpathlineto{\pgfqpoint{3.054473in}{1.181579in}}%
\pgfpathlineto{\pgfqpoint{3.161447in}{1.181925in}}%
\pgfpathlineto{\pgfqpoint{3.268422in}{1.182128in}}%
\pgfusepath{stroke}%
\end{pgfscope}%
\begin{pgfscope}%
\pgfpathrectangle{\pgfqpoint{0.505278in}{0.415000in}}{\pgfqpoint{2.894722in}{0.841837in}}%
\pgfusepath{clip}%
\pgfsetrectcap%
\pgfsetroundjoin%
\pgfsetlinewidth{1.505625pt}%
\definecolor{currentstroke}{rgb}{0.000000,0.000000,0.000000}%
\pgfsetstrokecolor{currentstroke}%
\pgfsetdash{}{0pt}%
\pgfpathmoveto{\pgfqpoint{0.701040in}{0.453265in}}%
\pgfpathlineto{\pgfqpoint{0.808015in}{0.453265in}}%
\pgfpathlineto{\pgfqpoint{0.914989in}{0.453265in}}%
\pgfpathlineto{\pgfqpoint{1.021963in}{0.453265in}}%
\pgfpathlineto{\pgfqpoint{1.128937in}{0.453265in}}%
\pgfpathlineto{\pgfqpoint{1.235912in}{0.453265in}}%
\pgfpathlineto{\pgfqpoint{1.342886in}{0.453265in}}%
\pgfpathlineto{\pgfqpoint{1.449860in}{0.453265in}}%
\pgfpathlineto{\pgfqpoint{1.556834in}{0.453265in}}%
\pgfpathlineto{\pgfqpoint{1.663808in}{0.453265in}}%
\pgfpathlineto{\pgfqpoint{1.770783in}{0.453265in}}%
\pgfpathlineto{\pgfqpoint{1.877757in}{0.453265in}}%
\pgfpathlineto{\pgfqpoint{1.984731in}{0.453265in}}%
\pgfpathlineto{\pgfqpoint{2.091705in}{0.453265in}}%
\pgfpathlineto{\pgfqpoint{2.198680in}{0.453265in}}%
\pgfpathlineto{\pgfqpoint{2.305654in}{0.453265in}}%
\pgfpathlineto{\pgfqpoint{2.412628in}{0.453265in}}%
\pgfpathlineto{\pgfqpoint{2.519602in}{0.453265in}}%
\pgfpathlineto{\pgfqpoint{2.626576in}{0.453265in}}%
\pgfpathlineto{\pgfqpoint{2.733551in}{0.453265in}}%
\pgfpathlineto{\pgfqpoint{2.840525in}{0.453265in}}%
\pgfpathlineto{\pgfqpoint{2.947499in}{0.453265in}}%
\pgfpathlineto{\pgfqpoint{3.054473in}{0.453265in}}%
\pgfpathlineto{\pgfqpoint{3.161447in}{0.453265in}}%
\pgfpathlineto{\pgfqpoint{3.268422in}{0.453265in}}%
\pgfusepath{stroke}%
\end{pgfscope}%
\begin{pgfscope}%
\pgfpathrectangle{\pgfqpoint{0.505278in}{0.415000in}}{\pgfqpoint{2.894722in}{0.841837in}}%
\pgfusepath{clip}%
\pgfsetbuttcap%
\pgfsetroundjoin%
\pgfsetlinewidth{1.505625pt}%
\definecolor{currentstroke}{rgb}{0.000000,0.500000,0.000000}%
\pgfsetstrokecolor{currentstroke}%
\pgfsetdash{{1.500000pt}{2.475000pt}}{0.000000pt}%
\pgfpathmoveto{\pgfqpoint{0.636856in}{0.453265in}}%
\pgfpathlineto{\pgfqpoint{0.636856in}{1.218571in}}%
\pgfusepath{stroke}%
\end{pgfscope}%
\begin{pgfscope}%
\pgfsetrectcap%
\pgfsetmiterjoin%
\pgfsetlinewidth{0.803000pt}%
\definecolor{currentstroke}{rgb}{0.000000,0.000000,0.000000}%
\pgfsetstrokecolor{currentstroke}%
\pgfsetdash{}{0pt}%
\pgfpathmoveto{\pgfqpoint{0.505278in}{0.415000in}}%
\pgfpathlineto{\pgfqpoint{0.505278in}{1.256836in}}%
\pgfusepath{stroke}%
\end{pgfscope}%
\begin{pgfscope}%
\pgfsetrectcap%
\pgfsetmiterjoin%
\pgfsetlinewidth{0.803000pt}%
\definecolor{currentstroke}{rgb}{0.000000,0.000000,0.000000}%
\pgfsetstrokecolor{currentstroke}%
\pgfsetdash{}{0pt}%
\pgfpathmoveto{\pgfqpoint{3.400000in}{0.415000in}}%
\pgfpathlineto{\pgfqpoint{3.400000in}{1.256836in}}%
\pgfusepath{stroke}%
\end{pgfscope}%
\begin{pgfscope}%
\pgfsetrectcap%
\pgfsetmiterjoin%
\pgfsetlinewidth{0.803000pt}%
\definecolor{currentstroke}{rgb}{0.000000,0.000000,0.000000}%
\pgfsetstrokecolor{currentstroke}%
\pgfsetdash{}{0pt}%
\pgfpathmoveto{\pgfqpoint{0.505278in}{0.415000in}}%
\pgfpathlineto{\pgfqpoint{3.400000in}{0.415000in}}%
\pgfusepath{stroke}%
\end{pgfscope}%
\begin{pgfscope}%
\pgfsetrectcap%
\pgfsetmiterjoin%
\pgfsetlinewidth{0.803000pt}%
\definecolor{currentstroke}{rgb}{0.000000,0.000000,0.000000}%
\pgfsetstrokecolor{currentstroke}%
\pgfsetdash{}{0pt}%
\pgfpathmoveto{\pgfqpoint{0.505278in}{1.256836in}}%
\pgfpathlineto{\pgfqpoint{3.400000in}{1.256836in}}%
\pgfusepath{stroke}%
\end{pgfscope}%
\begin{pgfscope}%
\pgfsetbuttcap%
\pgfsetmiterjoin%
\definecolor{currentfill}{rgb}{1.000000,1.000000,1.000000}%
\pgfsetfillcolor{currentfill}%
\pgfsetfillopacity{0.800000}%
\pgfsetlinewidth{1.003750pt}%
\definecolor{currentstroke}{rgb}{0.800000,0.800000,0.800000}%
\pgfsetstrokecolor{currentstroke}%
\pgfsetstrokeopacity{0.800000}%
\pgfsetdash{}{0pt}%
\pgfpathmoveto{\pgfqpoint{1.837083in}{0.594113in}}%
\pgfpathlineto{\pgfqpoint{3.302778in}{0.594113in}}%
\pgfpathquadraticcurveto{\pgfqpoint{3.330556in}{0.594113in}}{\pgfqpoint{3.330556in}{0.621890in}}%
\pgfpathlineto{\pgfqpoint{3.330556in}{1.049946in}}%
\pgfpathquadraticcurveto{\pgfqpoint{3.330556in}{1.077723in}}{\pgfqpoint{3.302778in}{1.077723in}}%
\pgfpathlineto{\pgfqpoint{1.837083in}{1.077723in}}%
\pgfpathquadraticcurveto{\pgfqpoint{1.809306in}{1.077723in}}{\pgfqpoint{1.809306in}{1.049946in}}%
\pgfpathlineto{\pgfqpoint{1.809306in}{0.621890in}}%
\pgfpathquadraticcurveto{\pgfqpoint{1.809306in}{0.594113in}}{\pgfqpoint{1.837083in}{0.594113in}}%
\pgfpathclose%
\pgfusepath{stroke,fill}%
\end{pgfscope}%
\begin{pgfscope}%
\pgfsetbuttcap%
\pgfsetroundjoin%
\pgfsetlinewidth{1.505625pt}%
\definecolor{currentstroke}{rgb}{0.121569,0.466667,0.705882}%
\pgfsetstrokecolor{currentstroke}%
\pgfsetdash{{5.550000pt}{2.400000pt}}{0.000000pt}%
\pgfpathmoveto{\pgfqpoint{1.864861in}{0.973557in}}%
\pgfpathlineto{\pgfqpoint{2.142639in}{0.973557in}}%
\pgfusepath{stroke}%
\end{pgfscope}%
\begin{pgfscope}%
\definecolor{textcolor}{rgb}{0.000000,0.000000,0.000000}%
\pgfsetstrokecolor{textcolor}%
\pgfsetfillcolor{textcolor}%
\pgftext[x=2.253750in,y=0.924946in,left,base]{\color{textcolor}\rmfamily\fontsize{10.000000}{12.000000}\selectfont Min Vehicle Dist}%
\end{pgfscope}%
\begin{pgfscope}%
\pgfsetrectcap%
\pgfsetroundjoin%
\pgfsetlinewidth{1.505625pt}%
\definecolor{currentstroke}{rgb}{0.000000,0.000000,0.000000}%
\pgfsetstrokecolor{currentstroke}%
\pgfsetdash{}{0pt}%
\pgfpathmoveto{\pgfqpoint{1.864861in}{0.849390in}}%
\pgfpathlineto{\pgfqpoint{2.142639in}{0.849390in}}%
\pgfusepath{stroke}%
\end{pgfscope}%
\begin{pgfscope}%
\definecolor{textcolor}{rgb}{0.000000,0.000000,0.000000}%
\pgfsetstrokecolor{textcolor}%
\pgfsetfillcolor{textcolor}%
\pgftext[x=2.253750in,y=0.800779in,left,base]{\color{textcolor}\rmfamily\fontsize{10.000000}{12.000000}\selectfont \(\displaystyle D_s = 5\)}%
\end{pgfscope}%
\begin{pgfscope}%
\pgfsetbuttcap%
\pgfsetroundjoin%
\pgfsetlinewidth{1.505625pt}%
\definecolor{currentstroke}{rgb}{0.000000,0.500000,0.000000}%
\pgfsetstrokecolor{currentstroke}%
\pgfsetdash{{1.500000pt}{2.475000pt}}{0.000000pt}%
\pgfpathmoveto{\pgfqpoint{1.864861in}{0.725224in}}%
\pgfpathlineto{\pgfqpoint{2.142639in}{0.725224in}}%
\pgfusepath{stroke}%
\end{pgfscope}%
\begin{pgfscope}%
\definecolor{textcolor}{rgb}{0.000000,0.000000,0.000000}%
\pgfsetstrokecolor{textcolor}%
\pgfsetfillcolor{textcolor}%
\pgftext[x=2.253750in,y=0.676613in,left,base]{\color{textcolor}\rmfamily\fontsize{10.000000}{12.000000}\selectfont Min Allowed \(\displaystyle R\)}%
\end{pgfscope}%
\end{pgfpicture}%
\makeatother%
\endgroup%

%% file: conclusion.tex
\section{CONCLUSION}

\label{sec_conclusion}

In this paper we have discussed
practical issues that arise when 
using a barrier function
to ensure safe operations when 
there is limited range sensing and actuator constraints.
The solution derives a new
barrier function that can be used to ensure that 
a system will stay safe for all future times
even though the system is still subject
to limited range sensing and was verified in a 20 UAV simulation study.